\documentstyle[epsf]{mn}

\newif\ifAMStwofonts

\def\kms{\relax \ifmmode {\,\rm km\,s}^{-1}\else \,km\,s$^{-1}$\fi}
\def\ha{\relax \ifmmode {\rm H}\alpha\else H$\alpha$\fi}
\def\hb{\relax \ifmmode {\rm H}\beta\else H$\beta$\fi}
\def\hi{\relax \ifmmode {\rm H\,{\sc i}}\else H\,{\sc i}\fi}
\def\hii{\relax \ifmmode {\rm H\,{\sc ii}}\else H\,{\sc ii}\fi}
\def\h2{\relax \ifmmode {\rm H}_2\else H$_2$\fi}
\def\lha{\relax \ifmmode L_{{\rm H}\alpha}\else $L_{{\rm H}\alpha}$\fi}
\def\shi{\relax \ifmmode \sigma_{{\rm HI}}\else $\sigma_{\rm HI}$\fi}
\def\sh2{\relax \ifmmode \sigma_{{\rm H}_2}\else $\sigma_{{\rm H}_2}$\fi}
\def\degr{\hbox{$^\circ$}}
\def\arcmin{\hbox{$^\prime$}}
\def\arcsec{\hbox{$^{\prime\prime}$}}
\def\deg{\hbox{$^\circ$}}
\def\min{\hbox{$^\prime$}}
\def\sec{\hbox{$^{\prime\prime}$}}
\def\fdg{\hbox{$.\!\!^\circ$}}
\def\fs{\hbox{$.\!\!^{\rm s}$}}
\def\farcm{\hbox{$.\mkern-4mu^\prime$}}
\def\farcs{\hbox{$.\!\!^{\prime\prime}$}}
\def\degd#1.#2{ #1\fdg#2 }                 
\def\mind#1.#2{ #1\farcm#2 }               
\def\secd#1.#2{ #1\farcs#2 }               
\def\hhh{\ifmmode {\rm ^h}              
         \else {${\rm ^h}$}
         \fi}
\def\sss{\ifmmode {\rm ^s}              
         \else {${\rm ^s}$}
         \fi}
\def\hms#1h#2m#3s{                      
                  \relax
                  \ifmmode #1^{\rm h}\,#2^{\rm m}\,#3^{\rm s}
                  \else \hbox{$#1^{\rm h}\,#2^{\rm m}\,#3^{\rm s}$}
                  \fi
                 }
\def\dms#1d#2m#3s{                      
                  \relax
                  #1\degr\,#2\arcmin\,#3\arcsec 
                 }
\def\hmsd#1h#2m#3.#4s{                  
                      \relax
                      \ifmmode #1^{\rm h}\,#2^{\rm m}\,#3\fs#4
                      \else \hbox{$#1^{\rm h}\,#2^{\rm m}\,#3\fs#4$}
                      \fi
                     }
\def\dmsd#1d#2m#3.#4s{                  
                      \relax
                      #1\degr\,#2\arcmin\,#3\farcs#4
                     }
\def\mag{\relax                          
        \ifmmode ^{\rm m}
        \else $^{\rm m}$
        \fi
       }
\def\magd#1.#2{                          
              \relax
              \ifmmode #1^{\rm m}
                       \hskip-0.55em.\hskip0.22em#2
              \else \hbox{#1$^{\rm m}
                    \hskip-0.55em.\hskip0.22em$#2}
              \fi
             }

\ifoldfss
  \ifCUPmtlplainloaded \else
    \NewTextAlphabet{textbfit} {cmbxti10} {}
    \NewTextAlphabet{textbfss} {cmssbx10} {}
    \NewMathAlphabet{mathbfit} {cmbxti10} {} 
    \NewMathAlphabet{mathbfss} {cmssbx10} {} 
  \fi
  \ifAMStwofonts
    \ifCUPmtlplainloaded \else
      \NewSymbolFont{upmath} {eurm10}
      \NewSymbolFont{AMSa} {msam10}
      \NewMathSymbol{\upi}     {0}{upmath}{19}
      \NewMathSymbol{\umu}     {0}{upmath}{16}
      \NewMathSymbol{\upartial}{0}{upmath}{40}
      \NewMathSymbol{\leqslant}{3}{AMSa}{36}
      \NewMathSymbol{\geqslant}{3}{AMSa}{3E}

    \fi
  \fi
\fi 

\ifnfssone
  \newmathalphabet{\mathit}
  \addtoversion{normal}{\mathit}{cmr}{m}{it}
  \addtoversion{bold}{\mathit}{cmr}{bx}{it}
  \newmathalphabet{\mathbfit} 
  \addtoversion{normal}{\mathbfit}{cmr}{bx}{it}
  \addtoversion{bold}{\mathbfit}{cmr}{bx}{it}
  \newmathalphabet{\mathbfss} 
  \addtoversion{normal}{\mathbfss}{cmss}{bx}{n}
  \addtoversion{bold}{\mathbfss}{cmss}{bx}{n}
  \ifAMStwofonts
    \ifCUPmtlplainloaded \else
      \UseAMStwoboldmath
      \makeatletter
      \new@mathgroup\upmath@group
      \define@mathgroup\mv@normal\upmath@group{eur}{m}{n}
      \define@mathgroup\mv@bold\upmath@group{eur}{b}{n}
      \edef\UPM{\hexnumber\upmath@group}
      \new@mathgroup\amsa@group
      \define@mathgroup\mv@normal\amsa@group{msa}{m}{n}
      \define@mathgroup\mv@bold\amsa@group{msa}{m}{n}
      \edef\AMSa{\hexnumber\amsa@group}
      \makeatother
      \mathchardef\upi="0\UPM19
      \mathchardef\umu="0\UPM16
      \mathchardef\upartial="0\UPM40
      \mathchardef\leqslant="3\AMSa36
      \mathchardef\geqslant="3\AMSa3E
    \fi
  \fi
\fi 

\ifnfsstwo
  \DeclareMathAlphabet{\mathbfit}{OT1}{cmr}{bx}{it}
  \SetMathAlphabet\mathbfit{bold}{OT1}{cmr}{bx}{it}
  \DeclareMathAlphabet{\mathbfss}{OT1}{cmss}{bx}{n}
  \SetMathAlphabet\mathbfss{bold}{OT1}{cmss}{bx}{n}
  \ifAMStwofonts
    \ifCUPmtlplainloaded \else
      \DeclareSymbolFont{UPM}{U}{eur}{m}{n}
      \SetSymbolFont{UPM}{bold}{U}{eur}{b}{n}
      \DeclareSymbolFont{AMSa}{U}{msa}{m}{n}
      \DeclareMathSymbol{\upi}{0}{UPM}{"19}
      \DeclareMathSymbol{\umu}{0}{UPM}{"16}
      \DeclareMathSymbol{\upartial}{0}{UPM}{"40}
      \DeclareMathSymbol{\leqslant}{3}{AMSa}{"36}
      \DeclareMathSymbol{\geqslant}{3}{AMSa}{"3E}
    \fi
  \fi
\fi 

\ifCUPmtlplainloaded \else
  \ifAMStwofonts \else 
    \def\upi{\pi}
    \def\umu{\mu}
    \def\upartial{\partial}
  \fi
\fi

\title[Stars, gas and dust in M100]{Global morphology and physical
relations between the stars, gas and dust in the disc and arms of M100}
\author[J. H. Knapen \& J.E. Beckman]
	{J. H. Knapen$^{1,2}$
        and J.E. Beckman$^3$\\ 
$^1$Department of Physical Sciences, Division of Physics and Astronomy, 
University of Hertfordshire, College Lane, Hatfield,\\
Herts AL10 9AB, UK. E-mail
knapen@star.herts.ac.uk\\
$^2$D\'epartement de Physique, Universit\'e de Montr\'eal, C.P.  6128,
Succursale Centre Ville, Montr\'eal (Qu\'ebec), H3C 3J7 Canada;\\ and
Observatoire du Mont M\'egantic\\
$^3$Instituto de Astrof\'\i sica de Canarias, E-38200 La Laguna,
Tenerife, Spain. E-mail jeb@iac.es }

\date{Accepted July 1996,
      Received;
      in original form}

\pagerange{\pageref{firstpage}--\pageref{lastpage}}
\pubyear{1995}

\begin{document}

\maketitle

\label{firstpage}

\begin{abstract}

We study star formation processes in the disc of the weakly barred grand
design spiral galaxy M100 (NGC~4321) from a variety of images tracing
recent massive star formation, old and young stars, dust, and neutral
hydrogen.  Differences between arm and interarm regions are specifically
studied by decomposing the images into arm and non-arm zones.  We find
from a comparison of the morphology in \ha, \hi\ and dust that while the
first two are coincident over most of the disc, they are offset from the
dust lanes especially along the inner parts of the spiral arms: a
picture which is indicative of a density wave shock moving through the
arms. \hi\ is formed near the young massive stars as a result of
photo-dissociation.

From radial profiles we find that in the region of the star-forming
spiral arms the exponential scale lengths for \ha, blue and
near-infrared light, and 21 cm radio continuum are equal within the
fitting errors.  The scale lengths for the interarm region are also
equal for all these tracers, but the arm scale lengths are
significantly longer.  This points to a common origin of the profiles
in star formation, with little or no influence from radial population
gradients or dust in the disc of this galaxy.  The longer arm scale
lengths are equivalent to an outwardly increasing arm-interarm
contrast. We argue that the radial profiles of radio continuum and
\hi, as well as CO, are also directly regulated by star formation, and
discuss the possible implications of this result for the
interpretation of observed CO intensities in and outside spiral arms.

We discuss the radial atomic hydrogen profile in some detail. Its almost
perfectly flat shape in the region of the star-forming spiral arms may
be explained by photodissociation and recombination processes in the
presence of a limited quantity of interstellar dust, controlling the
equilibrium between the molecular and atomic form of hydrogen.  Over
most of the inner part of the disc, H{\sc i} seems to be a {\it product}
of the star formation processes, rather than the {\it cause} of enhanced
star formation.

\end{abstract}


\section{Introduction}

In order to learn more about the interaction between the different
components of the ISM and about the processes leading to star formation
(SF) in the arms of spiral galaxies, it is important to compare directly
arm regions with those between the arms (interarm regions).  So far,
only a limited number of grand-design, late-type spiral galaxies has
been studied in detail in this sense combining information from several
tracers of SF and neutral gas throughout their discs.  This is because
the resolution attainable with radio and millimeter observations entails
the selection of galaxies with large angular size, which implies long
mapping times. For both interferometric observations of \hi\ and
single-dish CO measurements the limiting angular resolution is around
15\sec, corresponding to slightly over 1~kpc in the case of NGC~4321,
the relatively nearby Virgo spiral studied in the present paper. Given
that the scale of the spiral arms is typically also some 1~kpc, galaxies
at larger distances are not suitable for this kind of study.

Comparing arm with interarm regions, Cepa \& Beckman (1990) found for
NGC~628 and NGC~3992 that the \ha/\hi\ ratios were higher in the arms
than between them, and that the arm/interarm ratios of \ha/\hi\ vary
congruently along the two main arms. They interpreted the pattern in the
arm/interarm ratios as indicating the locations of the density wave
resonances, and were able to determine a pattern speed in that
way. Tacconi \& Young (1990) found an enhanced \ha\ to CO ratio in the
northeast arm of NGC~6946 as compared to the neighbouring interarm
region, implying triggering of the star formation in the arm if the CO
reliably traces the molecular gas mass, but the southwest arm of that
galaxy does not show such an enhanced \ha/CO ratio.

The galaxy M51 has received quite a lot of attention in this
respect. Lord \& Young (1990) studied arm and interarm regions
separately using CO and \ha\ data, but they were limited to the 45\sec\
resolution of their CO data which was not quite adequate to isolate the
spiral arms. Knapen et al. (1992) combined the Nobeyama CO map from
Nakai et al. (1991; see also Nakai et al. 1994) with \hi\ and \ha\ data
to determine the arm-interarm ratio of the massive star formation
efficiency, defined as \ha\ luminosity per unit total (atomic plus
molecular) gas. They found a pattern of dips and strong peaks in the
arm-interarm efficiency ratio, which is symmetric in the two main arms
of M51. The fact that the efficiency ratios are consistently larger than
unity along both arms is strong evidence for triggering of the massive
SF in the spiral arms. Garc\'\i a-Burillo, Gu\'elin \& Cernicharo
(1993), using IRAM CO data of slightly higher resolution than those used
by Knapen et al. (1992), confirmed the findings of the latter authors on
relative arm/interarm contrasts. For the galaxy studied in this paper,
NGC~4321, Knapen et al. (1996), using new CO data from Nobeyama (see
also Cepa et al. 1992) and combining it with \ha\ and \hi\ maps, found
that the massive SF efficiencies are enhanced in the arms with respect
to the interarm regions. The arm/interarm efficiency ratios are very
similar to those found by Knapen et al. (1992) in M51, but the symmetric
pattern of peaks and dips so obvious in M51 is absent from
NGC~4321. Knapen et al. (1996) argue that their conclusion on
triggering can only be strengthened if possible arm/interarm variations
in dust extinction or the CO to \h2\ conversion factor are considered.

In the present paper, we will discuss the spatial relationships
between not only CO, \hi, and \ha, but also blue and near-infrared
(NIR) light, and radio continuum. We will do this via direct
2-dimensional comparison of some tracers, and through a comparison of
radial (azimuthally averaged) profiles. Similar profiles for a number
of tracers have been discussed by (among others) Tacconi \& Young
(1986) for NGC~6946 and Rand, Kulkarni \& Rice (1992) for M51, but
nowhere in the literature has the specific difference between arm and
interarm environments been taken into account when studying radial
profiles. Apart from whole disc radial profiles, here we also study
radial profiles for the arm and interarm zones separately.

NGC~4321 (M100) is a weakly barred spiral galaxy with two main spiral
arms.  It hosts a circumnuclear starburst region in its central region,
which has been studied extensively by Knapen et al.  (1995a,b).  In this
paper, we will concentrate on a region between some 30\sec\ and 250\sec\
in radius (or $0.1(D_{25})/2<R<1.2(D_{25}/2)$) comprised of part of the
bar, and the disc of the galaxy.  The bar (Pierce 1986; Knapen et al. 
1993) extends to some 60\sec\ in radius.  We assume a distance to
NGC~4321 of 17.1 Mpc (Freedman et al.  1994), thus 12\sec\ corresponds
to 1~kpc. 

We define here as the ``north arm'' the arm originating at the end of
the bar east of the nucleus, passing north of the centre and continuing
toward the west and south-west. The arm we denominate as the ``south
arm'' starts west of the nucleus at the end of the bar, passes south of
the central region, and continues toward the east and north-east.

After a short description of the observational data used (Section~2), we
discuss the spatial correlation of radio and optical emission in
Sect.~3. In Sect.~4 we present radial profiles for the whole disc of the
galaxy, but also for arm and interarm zones separately, and compare
exponential scale lengths fitted to the radial distributions. We discuss
the results critically in terms of star formation processes in Sect.~5,
with a specific discussion of the origin and role of \hi\ in Sect.~6. We
briefly summarize the main conclusions in Sect.~7.

\section{Observational Material}

In the present paper, we use data on NGC~4321 from our own published
research papers.  In this Section, they will be reviewed only very
briefly, since more technical details can be found in the original
studies.  The radio data used are from Knapen et al.  (1993): an \hi\
map of the whole galaxy at 15\sec\ resolution obtained by moment
analysis from a uniformly weighted VLA data cube; and a 21cm radio
continuum image (also at 15\sec\ resolution), which is the map that was
subtracted from the individual channel maps in order to produce the \hi\
line data-set.  The \ha\ image of the disc of NGC~4321 was shown also by
Knapen et al.  (1993), and a detailed study of the statistics of the
\hii\ regions throughout the disc can be found in Knapen (1996).  Here
we start out with a slightly smoothed version of the original image
(with a resolution of 2\sec), but we will also use a further smoothed
version with resolution 15\sec, in order to compare directly with the
\hi\ data.  Pixel sizes of the original; 2\sec; and 15\sec\ resolution
images are \secd 0.{27;} \secd 0.{54;} and \secd2.{0,} respectively. 
The $B$ and $I$ images used here (also shown by Knapen et al.  1993) are
discussed in more detail by Beckman et al.  (1996).  A $B-I$ colour
index image was produced by dividing one image by the other.  Since the
central $\sim10$\sec\ part of the $I$ band image is saturated, the $B-I$
image is not reliable in this inner region.  A $B-I$ image shows both
effects due to changing stellar populations and localized extinction by
dust, two effects that can be relatively well separated morphologically
on a map, but less easily when studying integrated emission (such as in
radial profiles).  The total field of view of all the images used
contains the complete disc of NGC~4321, out to $>(D_{25}/2)$.  All
images were given a common orientation using the position of the centre
of the galaxy and those of stars in the field.  Finally, we refer to
single dish CO measurements of a number of points in the disc of
NGC~4321, and mostly along the inner parts of the two main spiral arms
(Cepa et al.  1992; Knapen et al.  1996). 

\section{Spatial (anti-)correlation of different tracers}

\begin{figure*}
  \epsfxsize=18cm \epsfbox{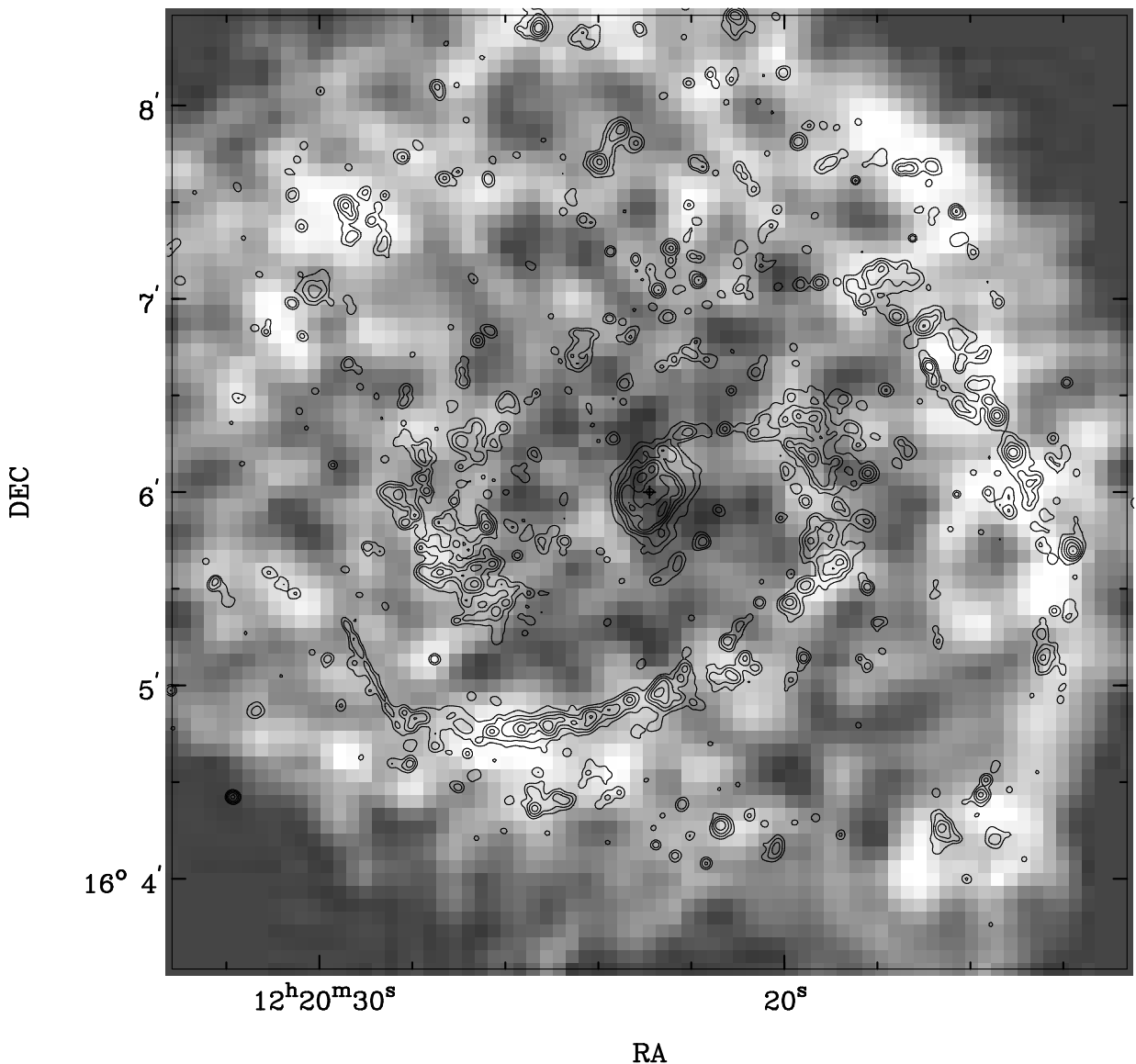}
  \caption{({\it a}) Overlay of the 2\sec\ resolution continuum-subtracted
\ha\ image of NGC~4321 (contours) on a grey-scale representation of the
\hi\ total intensity map at 15\sec\ resolution.  Contour levels are
0.37, 0.74, 1.48, 2.95, 5.90, 11.8, and $23.6\times10^{36}$
erg\,s$^{-1}$.  Greys are from 0.98 to $14.7\times10^{20}$
atoms\,cm$^{-2}$.}
\end{figure*}

\setcounter{figure}{0}

\begin{figure*}
  \epsfxsize=18cm \epsfbox{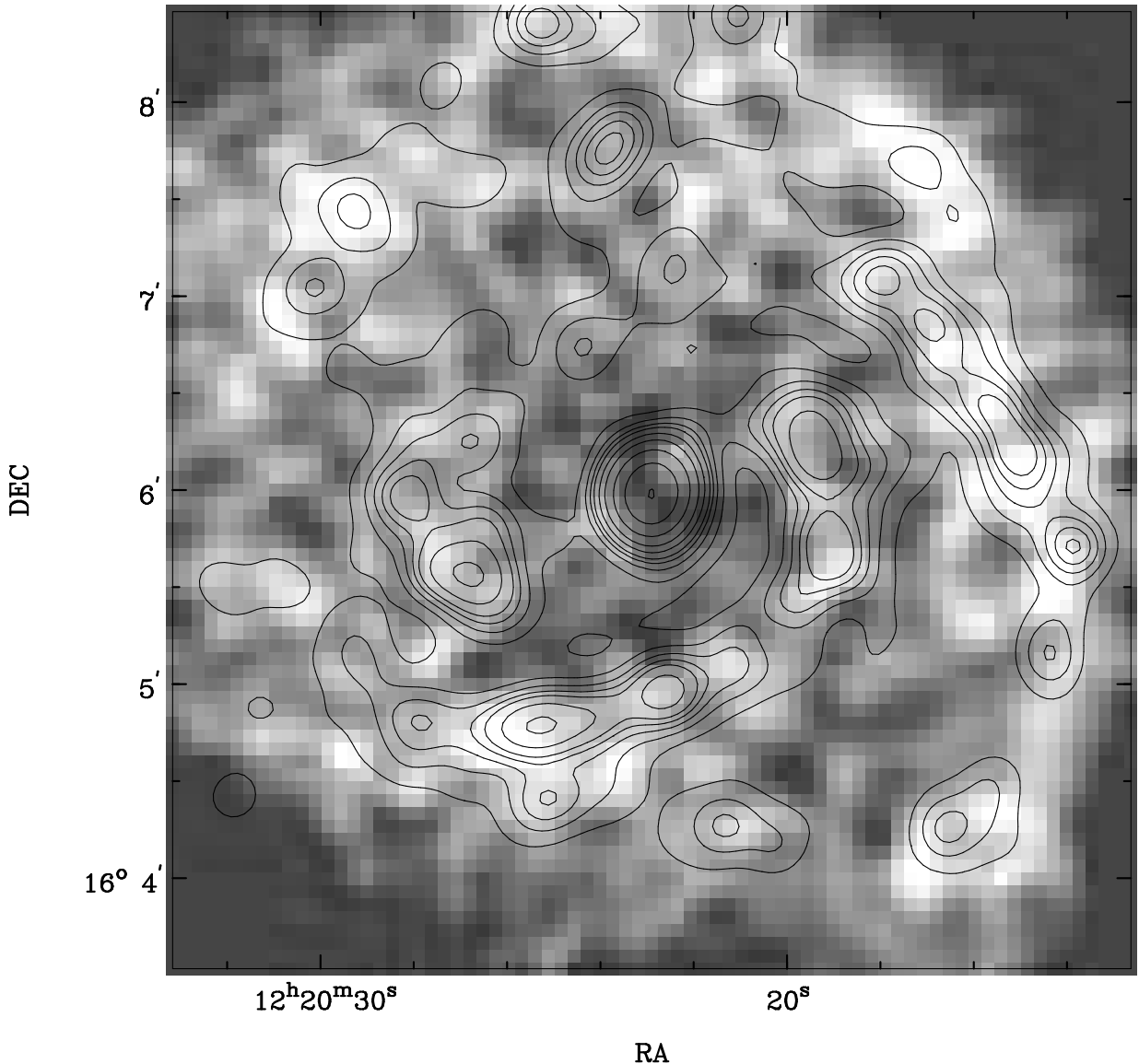}
  \caption{({\it b}) As Fig.~1a, but now a smoothed version (at 15\sec\
resolution) of the \ha\ image is shown overlaid. Contour levels are
0.12, 0.25, 0.37, 0.49, 0.62, 0.86, 1.23, 1.97, 3.20, and
$4.92\times10^{36}$ erg\,s$^{-1}$.}
\end{figure*}

\begin{figure*}
\vspace{12cm}
  \caption{The 2\sec\ \ha\ image from Fig.~1a shown overlaid on a
grey-scale representation of the $B-I$ colour image of the disc of
NGC~4321.  Redder colours (e.g.  dust lanes) are represented by darker
shades, bluer colours (e.g.  sites of recent or ongoing SF) by lighter
shades.  Greys range from 0.7 mag (white) to 2.4 mag (black).  Contours
as in Fig.~1. Note that the $B-I$ image cannot be used at those pixel
positions where the $I$ image was saturated (mostly in the nuclear
region). These pixel values were discarded and show up black in the
Figure.}
\end{figure*}

\begin{figure*}
\vspace{12cm}
\caption{The \hi\ image from Fig.~1, now in contour representation,
shown overlaid on the $B-I$ colour image of Fig.~2.  Greys as in Fig.~2. 
\hi\ contours are from 4.88 to 14.7 in steps of $1.95\times10^{20}$
atoms\, cm$^{-2}$.  Note that lower \hi\ contours have been emitted for
clarity from this Figure (compare with Knapen et al.  1993).}
\end{figure*}

In order to study the relative location of dust, young stars, and atomic
hydrogen, we present overlays of \ha\ (which traces young massive stars)
on \hi\ (atomic hydrogen) (Fig.~1 a,b), \ha\ on $B-I$ (indicating the
location of the dust lanes; Fig.~2), and \hi\ on $B-I$ (Fig.~3).  We
will compare these with overlays of interferometric CO observations
(Rand 1995) on some of the images also used in the present paper,
especially Rand's Fig.~2 showing CO on our \ha\ and $B-I$ in a selected
region W and S of the nucleus.

Fig~1a shows the \ha\ at high resolution overlaid on \hi, while in
Fig.~1b the smoothed \ha\ image (15\sec\ resolution) is overlaid on the
\hi\ image of the same resolution.  The morphology of the disc of
NGC~4321 is apparent from the \ha.  The strong \ha\ emission of the
central region is in fact organized in a two-armed mini-spiral (see
Knapen et al.  1995a,b for a detailed discussion).  The weak stellar bar
(Pierce 1986; Knapen et al.  1993) extends to some 40\sec\ in radius,
and does not show much \ha\ emission.  The main spiral arms are well
defined in \ha\ outside two strongly emitting regions at the ends of the
bar.  Although the general arm shape is that of a two-armed grand design
galaxy with two symmetric spiral arms, the \ha\ distribution along the
arms outside the SF regions at the ends of the bar, is in fact
anti-symmetric.  Along the south arm, \ha\ emission is strong in the
region parallel to and south of the bar, whereas there is considerably
less \ha\ emission along the north arm parallel to the bar (see Knapen
et al.  1996 for a discussion of SF and efficiencies along these arm
segments).  Further along the arms, the \ha\ emission picks up along the
north arm (near $\alpha=\hms 12h20m18s$, $\delta=16\deg7\min$) but
almost completely disappears in the other arm ($\alpha=\hms 12h20m28s$,
$\delta=16\deg5\min$).  The strong \ha\ emission along the north arm
stops near $R\sim120\sec$ (near $\alpha=\hms 12h20m15s$,
$\delta=16\deg6\min$), and outside that radius the \ha\ emission along
both arms is weak and not continuous, although still outlining the
spiral arms (see also the next section).  Interarm \ha\ emission is
present in large portions of the disc, but is usually weak, and arises
from small, isolated regions, which in some cases are however lined up
along spiral arm fragments.

\subsection{\hi\ and \ha}

From both Fig.~1a and b it is clear that along the two main spiral arms,
the \ha\ emission generally coincides with the \hi.  The \hi\ emission
is depressed in the circumnuclear and bar regions, and is not very
strong along the north arm segment parallel to the bar.  But especially
along the arm segments south and west of the nucleus, where the \ha\ is
strong and defines the arm shape, the \hi\ coincides well with the \ha. 
From Fig.~1b it is clear that most regions emitting strongly in \hi\
(white in the grey-scale Figure) are accompanied by a patch of \ha\
emission.  This is even more evident from the high resolution \ha\
overlay (Fig.~1a), which shows that practically all regions of \hi\
emission are accompanied by one or more \hii\ regions emitting in \ha. 
Since many small \hii\ regions are scattered over most of the disc
surface, it is important to note that most dark spots (no, or reduced,
\hi\ emission) in the figure are {\it not} accompanied by any \ha\
emission.  We thus find that along most of the spiral arms, and
certainly where the arms are well defined and strong in \ha, the \hi\
and \ha\ emission spatially coincide, but also that individual patches
of \hi\ emission over the whole disc surface are generally accompanied
by \ha\ emission. 


\subsection{Dust lanes}

Fig.~2 shows an overlay of the \ha\ image on the $B-I$ colour index map. 
Although in a $B-I$ image alone the detailed effects due to extinction
by dust and changing stellar populations (which can both result in a
redder colour) cannot be distinguished, the $B-I$ image does indicate
the location of the major dust lanes in the disc of the galaxy.  When
compared with the true-colour image of NGC~4321 presented by Beckman et
al.  (1996; see also Peletier 1994), it is evident that the dark lanes
along the arms and in the central (bar) and interarm regions are in fact
dust lanes, as is also indicated by their characteristic patchy
appearance. 

Fig.~2 shows that the \ha\ emission is generally offset from the dust
lanes, in the spiral arms, in the bar, and in some cases also in the
interarm region (see e.g.  the region east of the bar).  This picture is
indicative of the presence of a density wave pattern in which the dust
lanes indicate the gas compression zone, and the \ha\ results from young
stars being formed just downstream from there (the North side of the
galaxy is the approaching side [Knapen et al. 1993], thus the downstream
side of the arm in this region which is inside corotation [e.g. Knapen
et al. 1996] is the convex side).  This is seen especially clearly along
the first $\sim90$\deg\ in PA of the south arm, starting immediately
outside the circumnuclear zone, and also along most of the north arm.
The shape of the dust lanes in the bar zone is indicative of streaming
due to a weak bar (Athanassoula 1992); the lanes are accompanied by a
string of \hii\ regions (see also Knapen et al.  1995b).

Along the part of the south arm immediately south of the centre, the
situation is not quite as clear.  The dust lane bifurcates, and in fact
surrounds the \ha\ emission in this part of the arm, although the
stronger lane remains on the concave side of the \ha\ arm. In the outer
part of the disc ($R>120$\sec), where the arms are not well defined in
\ha, dust lanes are hardly visible and the situation is much less clear.

Fig.~3 shows graphically what can in principle be deduced from Figs.~1
and 2: that the \hi\ is generally offset from the dust lanes, and
coincides with the SF regions, which have blue colours in $B-I$. The
general offset between dust lanes and \hi\ spiral arms is particularly
clear from this overlay. Many of the regions where \hi\ and \ha\
emission coincide, as discussed above from Fig.~1, can be recognized.

\subsection{CO}

Rand (1995) presented overlays of his new BIMA interferometric CO image
on the same \ha\ and $B-I$ images as used in the present paper.  His CO
data cover the south arm only up to where the strong \ha\ complex ends. 
Rand finds that in the bar and along the first part of the arm, the CO
coincides with the dust lanes, and is offset from the \ha.  He
interprets the behaviour in the bar and along the inner part of the arm
as an indication that the star formation is triggered by gas compression
within the bar potential, and by the density wave, respectively. 
Further along the arm, where the dust lane bifurcates, the CO emission
coincides with the \ha, and not with the dust lanes.  According to Rand,
this zone may be close to corotation (where no offset is expected
between stars and the dense gas from which they form), but the dust
lanes may trace diffuse gas that has little to do with SF.  An
alternative explanation would be that the CO does not trace the
molecular hydrogen, but is in fact seen where it is heated by the SF
activity (see below).  Comparison of the relative locations of this arm
segment in CO and (non-thermal) radio continuum could shed light on the
importance of cosmic-ray heating of the CO (cf.  Adler, Allen \& Lo
1991; Allen 1992), but unfortunately the 21~cm map used here is not of
sufficient quality to use in this sense. 

\section{Radial Profiles}

\subsection{Whole-disc profiles}

\begin{figure}
  \epsfxsize=8cm \epsfbox{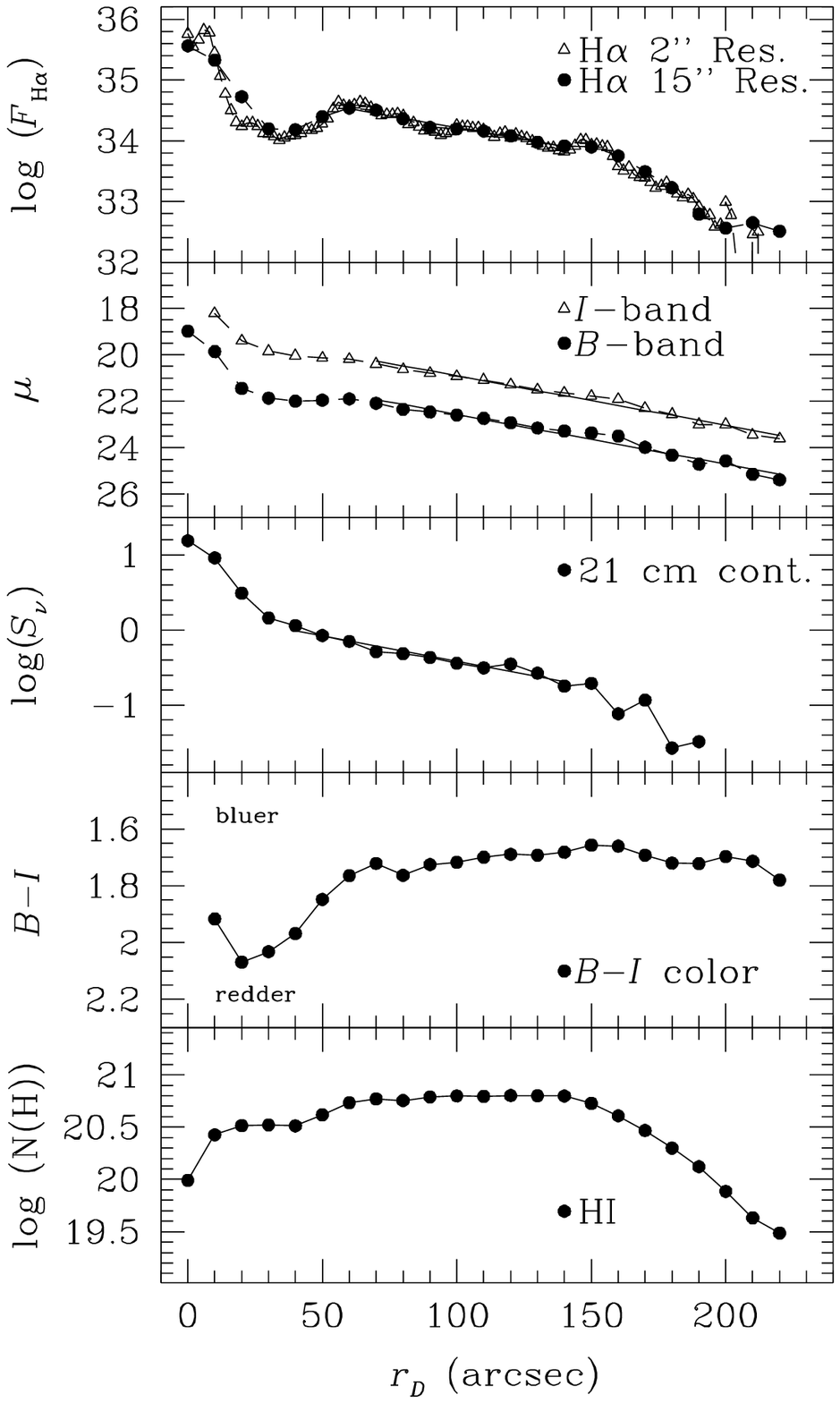}
  \caption{Azimuthally averaged radial profiles of \ha\ at 2\sec\ (open
  triangles) and 15\sec\ (filled dots) 
  resolution (upper panel); $B$ (blue; filled dots) and
  $I$ bands (NIR; open triangles) (second panel); 21 cm radio
  continuum (third panel); $B-I$ colour (fourth panel); and 21 cm \hi\
  (lower panel). Drawn lines indicate exponential fits to the data
  points; fits were made in the ranges where the lines are drawn. Units
  are log of erg\,s$^{-1}$\,arcsec$^{-2}$ for \ha, magnitudes for $B$ and
  $I$ bands, log of mJy/beam for 21cm continuum, magnitudes for $B-I$, and
  log of atoms\,${\rm cm}^{-2}\,{\rm arcsec}^{-2}$ for \hi.}
\end{figure}

Radial profiles of several gas and stellar tracers are shown in Fig.~4. 
All curves are azimuthally averaged radial profiles, made by integrating
in elliptical annuli, all centred on the nucleus of the galaxy.  We
assumed values of $i=27$\deg\ for the inclination angle of the galaxy,
and PA$=153$\deg\ for the position angle of the major axis (Knapen et
al.  1993); these values were kept constant for all ellipses.  Except
for the high-resolution \ha\ profile for which we used 2\sec\ ellipses
and spacings, all profiles were calculated integrating in 10\sec\ wide
ellipses, with radii also increasing by 10\sec\ per step.  In \ha\ we
determined profiles at two different resolutions: one with the same
resolution as the radio data (15\sec, from a smoothed \ha\ image), the
other at 2\sec resolution.  Apart from the inner region ($r_D<30\sec$)
there are no differences between the two \ha\ profiles which might
influence the interpretation or the fitting of scale lengths, so we feel
confident in using profiles determined from images (not only in \ha) at
resolutions of around 15\sec\ for the analysis below. 

Other profiles plotted in Fig.~4 represent the distributions of blue and
red light ($B$ and $I$ bands); 21 cm radio continuum; $B-I$ colour
index; and 21 cm \hi.  Note that all quantities are plotted either on a
magnitude or purely logarithmic scale, so that exponential profiles
appear as straight lines in the figure.

Some of the most important characteristics of the morphology of NGC~4321
can be recognized in the radial profiles, most easily so in the \ha.
The nuclear region extends to about 15\sec\ and emits strongly in \ha\
(Knapen et al.  1995a,b).  The radial \ha\ profile is sharply peaked
here ($0\sec<r_D<15\sec$).  The region of the (weak) bar, with low \ha\
emission, extents from some 15\sec\ to 40\sec\ in radius.  Two regions
of enhanced SF at the ends of the bar show up as a bump around
$r_D=60$\sec, followed by a region where the \ha\ luminosity falls off
exponentially, which is where the SF spiral arms in the disc of NGC~4321
are found.  After a break in the profile at $r_D\sim140$\sec, the
profile falls off rapidly.  The \ha\ disc ends near $r_D=200$\sec, or
$0.95\,(D_{25}/2)$.

The specific zones seen in the \ha\ profile can be recognized in the $B$
and $I$ profiles: a central peak, a depression in the bar region (less
pronounced in $I$, as expected for an older, redder, bar population),
and enhanced emission from the region of the SF spiral arms (though less
enhanced than in \ha).  The radio continuum profile also shows the
central peak, followed by an exponential decline, not very different
from the optical profiles.  The $B-I$ colour profile is notable in that
it is practically constant, after a blue peak due to the SF in the
central region and a red depression in the bar zone.  The \hi\ profile
is the only one not showing a central peak, and in fact the main
resemblance it shows to the other profiles is that it is slightly
enhanced between $60\sec<r_D<140\sec$, or exactly in the region defined
above as the SF spiral arm domain.

\begin{table}
  \caption{Exponential scale lengths ($h$, in arcsec) for different
tracers (\ha\ at 15\sec\ resolution, $B$, $I$ and 21cm continuum) for
the disc of NGC~4321, as well as for the arm and interarm zones
separately.}
\begin{tabular}{llll}
\hline\hline
$h$ (\sec)	& Disc		& Arm		& Interarm     \\ 
\hline
\ha  		& 55.4$\pm$3	& 85.9$\pm$3	& 61.1$\pm$3 \\
$B$ 		& 50.7$\pm$3	& 79.3$\pm$3	& 55.5$\pm$3 \\
$I$  		& 50.9$\pm$3	& 77.6$\pm$3	& 54.3$\pm$3 \\
21cm cont. 	& 63.7$\pm$6	& 81.4$\pm$7	& 60.1$\pm$6 \\
\hline\hline
\end{tabular}
\end{table}

We fitted exponential scale lengths $h$ to the radial profiles in the
region of the SF spiral arms ($60\sec<r_D<150\sec$) by least squares
fits to the data points.  The results and their estimated uncertainties
are listed in Table~1, and the fits are indicated as continuous lines in
Fig.~4, where the extent of the line shows the exact range in radius
used for fitting in each case.  The main result is that scale lengths
for \ha, $B$ and $I$ do not differ significantly, whereas the radio
continuum scale length is marginally longer than the optical one, but is
within 1$\sigma$ of $h_{\ha}$.  Note that larger scale lengths in radio
continuum could be expected because of cosmic ray propagation (e.g. 
Helou \& Bicay 1993).  The \ha\ and optical scale lengths are also very
similar to the scale length determined for the CO
distribution\footnote{Note that the CO scale length was not obtained
from an azimuthally averaged radial profile, but from observed points in
the disc, which may be biased toward points on the spiral arms.  In
using the CO scale length as we have done here, one assumes that no
changes with radius of azimuthal variations in CO emission are present. 
The large error bars on $h_{\rm CO}$ reflect this.} ($h_{\rm
CO}=46\pm8\sec$, Kenney \& Young 1988; Knapen et al.  1996).  The
optical values agree with Grosb\o l's (1985) value of \secd 48.9, but
differ from the values found by Elmegreen \& Elmegreen (1984) of
$h_B=\secd71.2$\ and $h_I=\secd69.1$.  Such a difference may be due to
different ranges used for the fits (see Knapen \& van der Kruit 1991). 
Note that Beckman et al.  (1996) find slightly different values for the
scale lengths but these differences may well be due to the fact that
they did not fix the position angle and inclination in the fits, along
with different definition of arms and interarm regions.  Ryder \& Dopita
(1994) found that the \ha\ scale lengths for a sample of 34 nearby
southern spirals were in general larger (but smaller or equal in about
1/5 of the cases) than those in $V$ and $I$.  Where the \ha\ scale
lengths are longer, they find that the $V$ scale lengths are usually
also longer than those in $I$.  However, NGC~4321 was not included in
their sample, and since no unique general rule about the \ha\ versus $V$
scale length behaviour can be inferred from their work, it is hard to
comment further.  We conclude that, within the uncertainties, disc scale
lengths for NGC~4321 in \ha, $B$, $I$, 21cm continuum and CO are equal,
with $h_{\rm d}=53\sec\pm2\sec$. 

\subsection{Separate arm and interarm radial profiles}

\begin{figure*}
  \epsfxsize=18cm \epsfbox{hamask.pps}
  \caption{Outlines of the mask used to separate arm from interarm
regions (heavy contour) overlaid on a contour representation of the
2\sec\ resolution \ha\ image of Fig.~1a. \ha\ contours as in Fig.~1a.}
\end{figure*}

\begin{figure}
  \epsfxsize=8cm \epsfbox{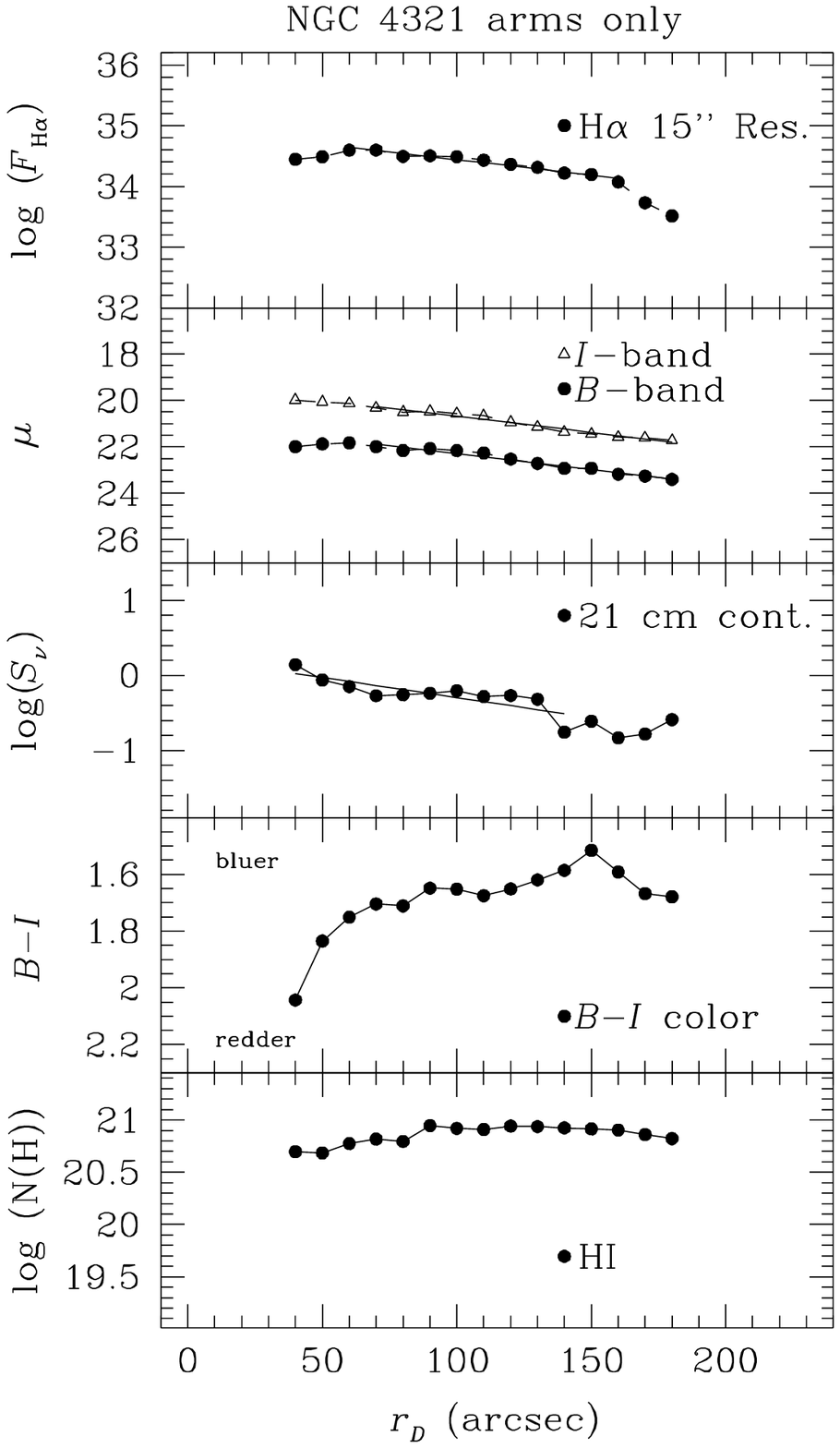}
  \caption{As Fig.~4, now for arms (a.) and interarm regions (b.)
separately.}
\end{figure}

\setcounter{figure}{5}

\begin{figure}
  \epsfxsize=8cm \epsfbox{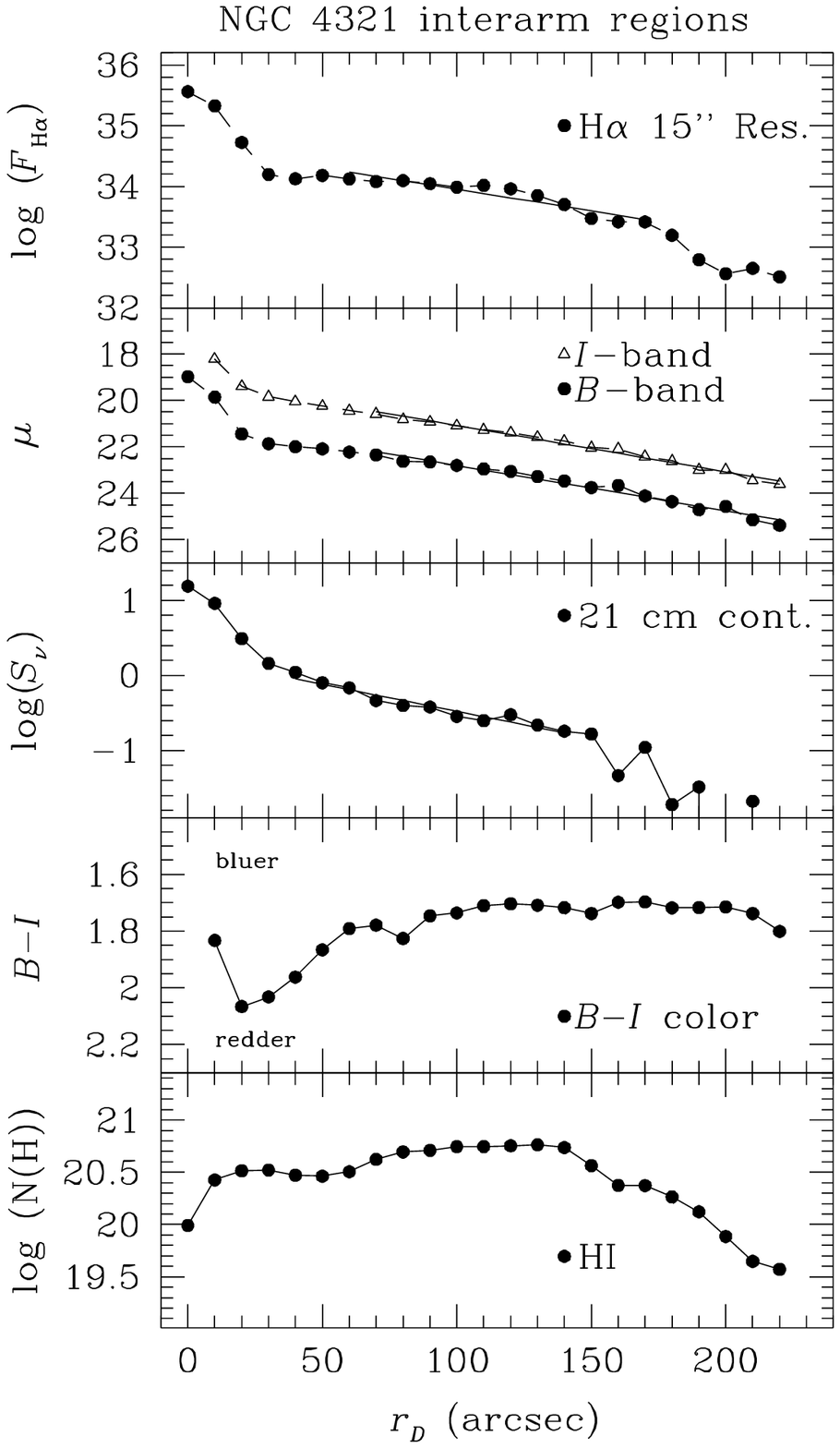}
  \caption{({\it b})}
\end{figure}

The similar scale lengths for optical, \ha, radio continuum and CO
emission point strongly toward a common origin.  To test this in more
detail, we produced separate arm and interarm images for all tracers,
except CO where no complete map of NGC~4321 is available.  The \ha\
image was used to identify the spiral arms, which were outlined
interactively.  A mask image was produced that was combined with the
\ha, $B$, $I$, radio continuum, $B-I$ and \hi\ images to make the
separate arm and interarm maps for all tracers.  The outline of the
adopted spiral arm region is shown overlaid on the \ha\ image in Fig.~5.
Following the ellipse fitting procedure described above for the whole
disc profiles (hereafter ``disc profiles''), we produced separate arm
and interarm radial profiles and fitted exponential scale lengths
(Fig.~6a,b; Table~1).


The interarm profiles and scale lengths look similar to those found for
the disc, which reflects the fact that most of the disc area is in the
interarm regions, whereas only a relatively small area can be defined as
arms.  Interarm scale lengths in \ha, $B$, $I$ and 21cm continuum are
equal within the uncertainties of the fit, even though in most cases
they are somewhat larger than those for the disc ($h_{\rm
ia}=56\sec\pm3\sec)$.  Although counterintuitive, the disc scale length
can be somewhat smaller than both the interarm and arm scale lengths due
to the outward decrease of the relative number of arm pixels. The arm
profiles are in all cases flatter than the disc and interarm profiles,
except for $B-I$ which is slightly rising.  The arm scale length values
reflect the flat profiles: they are more than 50\% larger than those for
the disc.  It is remarkable that arm scale lengths in \ha, $B$, $I$ and
radio continuum are again practically the same, around $h_{\rm
a}=81\sec\pm3\sec$.

The longer arm scale length can be interpreted as a direct effect of
enhanced SF in the arms, and is equivalent to the observation that the
arm-interarm contrast rises radially outward (where this contrast is
defined as the ratio of average arm intensity to average interarm
intensity).  Similar results were found earlier, using different
techniques, by Schweizer (1976) for 6 spiral galaxies, and by Carignan
(1985) who considered blue and red pixels separately for 3 late-type
spirals.  The fact that the values for optical, \ha\ and radio
continuum emission (and CO for the disc) are equal in NGC~4321 implies
that all these tracers are directly related to the SF activity.  We
will discuss these results, their validity and their implications in
more detail below.

\section{Star formation in discs of spiral galaxies}

In Sect.~4.1 we saw that the radial fall-off in the disc of NGC~4321 is
very similar for \ha, $B$, $I$, radio continuum, and CO, with equal
exponential scale lengths for all tracers.  The \hi\ and the $B-I$
colour profiles show a distinctly different behavior.  A similar picture
is seen in the galaxies NGC~6946 (Tacconi \& Young 1986) and M51 (Rand
et al. 1992), where the \ha\ and CO (and in the first case also blue
light and radio continuum) decline together, whereas the \hi\ shows a
depression in the centre, and joins the other profiles only in the outer
parts of the disc.

In the present paper we have gone one step further and decomposed the
profiles into arm and interarm contributions.  Whereas the scale lengths
in the arms and those in the interarm regions are significantly
different, the separate tracers: \ha, $B$, $I$, and radio continuum,
still have one common scale length in the arms, and another in the
interarm regions.  We will now discuss these results considering several
mechanisms that can affect them. 

\subsection{Dust}

First we must consider the effect of dust extinction and absorption on
the scale lengths as determined in the optical.  Assuming that the
amount of dust in a galactic disc declines radially (which happens
naturally if dust and stars are uniformly mixed), this will lead to
increasingly longer scale lengths toward the blue (Evans 1994; de Jong
1995; Peletier et al.  1995; Beckman et al.  1996). This is in fact
observed to varying degrees in other galaxies studied, indicating
varying amounts of dust in the disc. But in NGC~4321 we find equal
scale lengths in blue and NIR, although the arm scale lengths are
longer than the ones in the interarm regions. For NGC~4321, Beckman et
al.  (1996) claim that most of the disc is largely dust free, and
explain the seemingly contradictory observations of many
well-developed dust lanes in optical and colour images by the
confinement of the dust to these relatively narrow lanes so the effect
on the disc as a whole is small. Note that according to Beckman et
al. (1996) the other two galaxies they studied, M51 and NGC~3631, and
for which they did find longer scale lengths toward shorter
wavelengths, do contain significant quantities of dust distributed in
the disc.

The considerations of Beckman et al., and the fact that the scale
lengths in blue and NIR light, but also in \ha\ and radio continuum,
behave so similarly must lead to the conclusion that the disc of
NGC~4321 is indeed largely dust-free, and that the difference between
arm and interarm profiles cannot be a result of arm-interarm variations
in dust extinction.  The dust visible as the dust lanes, e.g.  in
Fig.~2, is mostly confined to those dust lanes, with the \ha\ emitting
regions aligned with the dust lanes but always offset from them.  The
fact that the $B$ and $I$ profiles are so similar in the disc (also
directly indicated by the nearly flat $B-I$ profile), and separately in
the arms and interarm regions as well, indicates that radial population
gradients must be very small.  If such gradients do exist, dust
extinction must exactly cancel their effects on the radial
profiles. This would amount to a conspiracy which we prefer to reject.

In fact we can go some way to setting quantitative limits on the dust
content if we assume that all the components of the arms, or of the
interarm disc, have the same geometrical (exponential) radial
distribution.  Taking this assumption, a formula by Regan \& Vogel
(1994) can be used to relate the on-axis, or equivalent face-on,
extinction at a given wavelength to the ratio of the measured scale
length at that wavelength and the true scale length: the formula
referred to is their equation (12).  We take the maximum permitted
differences between the measured $B$ and $I$ scale lengths, using the
errors quoted in Sect.~4, to obtain upper limits to the effective mean
dust content.  In both the arms and the interarm disc the on-axis dust
extinction estimated in this way must be less than 0.1 mag.  However,
as Beckman et al.  (1996) point out, an extinction of 0.1 mag obtained
this way can be reproduced by a dust lane of on-axis optical depth 1
but covering only 10\% of the system (arm or interarm disc) as seen
face-on.  In fact the presence of dust lanes or patches with much
greater optical depths could be missed in the above estimates,
provided that they cover only a moderate fraction of the face-on disc.
Thus our conclusion here is that the effective global extinction over
the arms or the interarm disc, must be low, but the present results do
not preclude the presence of non-uniform dust, compressed either in
lanes or into the plane of the galaxy, or both.

\subsection{Star Formation}

The four main sections of the \ha\ profile (central peak, dip in bar
region, exponential decline, and steeper decline after $R=160\sec$)
can be recognized in the \ha\ image (Fig.~1).  Enhanced SF in the
central region and in the main spiral arms between $50\sec<R<160\sec$
causes the relatively enhanced region in the radial profile, whereas
the lack of SF in the bar region coincides with the depression in the
radial profile.  The part of the profile where we determined the
exponential scale lengths coincides with the region where the
star-forming spiral arms are found, and it is tempting to conclude
that not only the \ha\ profile is controlled by SF, but also, since
the scale lengths are equal, the $B$ and $I$, radio continuum and CO
profiles (like in NGC~6946, Tacconi \& Young 1986).

This is confirmed by considering the separate arm-interarm profiles.
Having noted (in the previous Section) that dust extinction is not
varying significantly between the arms and the interarm regions, the
most important ground for difference between the arm and interarm
results is in the rate of SF.  If indeed the SF activity is the
underlying cause for the equal scale lengths for different tracers in
the disc as a whole, one might expect differences in arm and interarm
behaviour, congruent in all tracers, and showing more SF in the arms.
This is exactly what the disc profiles, as well as the separate arm
and interarm profiles show.  We conclude that the differences between
arms and interarm scale lengths, as well as the equal scale lengths
for \ha, $B$, $I$ and 21cm continuum in both the arm and interarm
regions separately, are due exclusively to enhanced SF activity in the
arms, and not (for example) to differential dust extinction.  This is
in accord with the conclusions reached by Knapen et al.  (1996), who
find that the massive SF rates but also efficiencies are at least some
2 or 3 times higher\footnote{This is a lower limit to the true
efficiency ratio if the CO intensity is controlled not only by the
quantity of molecular hydrogen but also by SF through e.g. enhanced
FUV flux, metallicity or cosmic-ray density, resulting in more
``hidden'' neutral gas in the interarm regions.} in the arms compared
to the regions outside the arms. Since much of the $I$-band light may
well be from the old disc, and not from recent SF activity, this would
imply also that the SF rate has not changed overall in the disc over a
substantial timescale. This would not be unusual, since in many
galaxies old and young discs have about the same intrinsic scale
length, apparent differences being due to dust (e.g. Peletier et
al. 1995; Regan \& Vogel 1995).

\subsection{Radio continuum and CO}

Young \& Scoville (1982) found that in NGC~6946 the CO radial profile
followed the blue light distribution. Combining Young \& Scoville's data
with \ha\ imaging, DeGioia-Eastwood et al. (1984) suggested that the SF
efficiency (SF rate per unit \h2\ surface density) is constant with
radius. For NGC~4321, we also find that the radial CO distribution
follows those of blue light and \ha. If the CO intensity is proportional
to the molecular gas density, this indeed means that the massive SF
efficiency does not vary radially in the disc.


A 21~cm radio continuum map, such as the one used in the present
paper, reflects a combination of thermal emission (mostly from ionized
gas in \hii\ regions) and non-thermal synchrotron emission (from
relativistic electrons in supernova remnants, and in the galactic
magnetic field). Although we lack spectral index information needed to
separate thermal from non-thermal emission, it is interesting to see
that the radial profiles for the disc, but also for the arm and
interarm zones, have the same scale lengths as the optical tracers. In
analogy with M51, we may expect that also in M100 the non-thermal
component is most likely to be stronger than the thermal one (Tilanus
et al. 1988 find that in M51 only some 5\% of the radio emission at
20~cm is thermal).

Since in the case of NGC~4321 both \hii\ regions (observed via \ha\
emission), young stars (via $B$) and old stars ($I$) behave very
similarly radially, it is not surprising that also the radio continuum
shows similar scale lengths in the radial distribution. This does,
however, confirm our conclusion that dust extinction cannot play an
important role in the disc of NGC~4321. If it did, blue light and \ha\
emission should be especially affected, but the radio continuum would
remain unchanged. Again, a special distribution of stars might
counteract the effects of dust, but this precise numerical cancellation
would be a ``conspiracy''.

Whereas a discussion of the detailed relation between stellar emission,
the ISM, and the non-thermal radio continuum is outside the scope of
this paper, we remark that the radio continuum is some measure of the
cosmic-ray density throughout the disc.  This has been discussed in some
detail by Adler et al.  (1991) and Allen (1992), who find that the ratio
of velocity-integrated CO emission and non-thermal radio continuum
brightness shows a remarkably small range within some large spiral
galaxies, but also among them.  This could imply that the CO intensity
is controlled by the UV flux and the cosmic ray density, more than by
the column density of molecular hydrogen (Allen 1992), as also discussed
by Rubio, Lequeux \& Boulanger (1993), Lequeux et al.  (1994), and Allen
et al.  (1995).  This, in turn, would imply that the observed radial CO
profile is the result of the distribution of the SF, rather than lying
at the origin of the latter.  This is an interesting idea that we cannot
pursue further with the present data, but which warrants further
research. 

\section{Origin and role of HI}

The radial \hi\ profile is perhaps the most challenging to our
understanding of all profiles presented here. In contrast to all other
tracers, a central peak is absent, corresponding to the central
depression seen on an \hi\ image of M100 (see Fig.~1 of the present
paper; also Warmels 1988; Cayatte et al. 1990; Knapen et al.  1993).
A central \hi\ depression is not at all uncommon in barred galaxies
(e.g. NGC~3992: Gottesman et al. 1984; NGC~1365: Ondrechen \& van der
Hulst 1989; J\"ors\"ater \& van Moorsel 1995), but some barred
galaxies do contain \hi\ in their central regions (e.g. NGC~5383:
Sancisi, Allen \& Sullivan 1979; NGC~4731: Gottesman et al 1984;
NGC~1097: Ondrechen, van der Hulst \& Hummel 1989). Of the 9
non-barred galaxies studied by Wevers, van der Kruit \& Allen (1986),
5 show evidence for a central \hi\ depression, whereas of the 6
galaxies classified as barred, 5 have a central depression. From the
shape of the radial \hi\ profiles, it is not possible to predict the
classification of the galaxy as barred or non-barred.  Broeils \& van
Woerden (1994) present radial \hi\ profiles of 48 spiral galaxies, of
which 14 are classified as barred. Some of the radial profiles derived
show a depression in the central region, others do not, and while
there is a trend with morphological type (with early-type spirals
having more of a central \hi\ depression), no systematic differences
in behaviour can be seen between barred and non-barred galaxies. We
thus conclude that the idea that bar dynamics cleans the bar region
from \hi\ by channeling the gas inward toward the nuclear zone and
outward toward the ``cusps'' at the ends of the bar, can in no case be
a general scenario. It is not solely the presence of a bar which
determines whether or not the \hi\ is depressed in the central
regions. Note that in some of these cases a bar in a galaxy may only
appear in NIR imaging and not in the optical, although it seems
unlikely that few-kpc length bars can be present in a substantial
fraction of galaxies which were not classified as SX or SB by e.g.  de
Vaucouleurs et al. (1991).  Another note to make here is that a
central depression in \hi\ cannot in general be taken as proof of a
lack of neutral gas.  A centrally peaked CO profile, such as seen in
NGC~4321, seems to be the standard, even for barred spirals with
\hi-poor central regions (e.g. Sandqvist et al. 1995; see also the
review by Young \& Scoville 1991).  As we will argue below, a central
atomic gas depression may well indicate changes in the equilibrium
between \hi\ and \h2.

The \hi\ disc profile of NGC~4321 shows an interesting plateau of
slightly enhanced emission, coinciding with the region of the SF spiral
arms as defined in \ha.  This indicates a relation between \hi\ and SF
activity, with either more SF occurring because of the enhanced atomic
gas density, or more \hi\ present because of the SF activity.  The first
possibility is contradicted by the CO measurements, which indicate that
molecular hydrogen is more abundant over the whole region, by more than
an order of magnitude up to $R\sim100\sec$ (Kenney \& Young 1988; Cepa
et al.  1992; Knapen et al.  1996).  We should note that there are no
definite indications so far that the CO to \h2\ conversion factor $X$ in
NGC~4321 differs from the assumed ``standard'' value measured in the
Galaxy (Rand 1995 found only weak, but inconclusive, evidence for a
lower factor), but that even if there is a variation of order 2-3, as
may well be the case in M51 (Adler et al. 1992; Rand 1993; Nakai \& Kuno
1995), it is extremely improbable that the \hi\ is more abundant than
\h2 in most of the region considered here.  Fig.~4 thus shows that \hi\
is more abundant where SF is present, and its increased column density
is almost certainly a result of the SF.  The shape of the radial \hi\
disc profile is very similar to the interarm profile (Fig.~6b), whereas
the \hi\ profile is almost completely flat in the arm regions (Fig.~6a).

A probable hypothesis for the origin of \hi\ is its production by
photodissociation of the molecular hydrogen by the UV radiation field of
young massive stars. Allen, Atherton \& Tilanus (1986) showed evidence
for this in M83 from the displacement of gas and dust from massive star
formation and \hi\ (see also Tilanus \& Allen 1989, 1991). Similar
offsets between tracers were also observed in M51 (Vogel et al. 1988;
Rand \& Kulkarni 1990). 

In M51, Knapen et al.  (1992) and Rand et al.  (1992) found that several
peaks in \ha\ were accompanied by peaks in \hi, and argued that the \hi\
peaks were caused by the enhanced SF activity through photodissociation
of molecular gas.  Again, the atomic hydrogen is almost certainly not
the cause of the star formation since the molecular hydrogen is an order
of magnitude more abundant at the radii where the peaks occur. 

In NGC~4321, evidence for the same process is seen using two separate
observational methods.  First, spatial correspondence is observed
between \hi\ and \ha, whereas both are clearly offset from the dust
lanes (Sect.~3).  The CO spiral arms are coincident with the dust
lanes in the inner part of the disc (Rand 1995).  This situation is
very similar to that observed in M51 (see above), but the situation in
the arm region just south of the centre is not clear (see Sect.~3.3),
and more reminiscent of that in M83 (Lord \& Kenney 1991).  Second,
radial profiles (Sect.~4) show that although the \hi\ behaves very
differently from other tracers, which are all directly related to SF
(see before), it is enhanced in the region where the spiral arms are
forming stars most actively.

The globally flat \hi\ column density profile over a wide radial range
in NGC~4321 is a feature relatively common in spirals.  An explanation
was given by Shaya \& Federman (1987) who suggested that the \hi\ column
density observed corresponds to that required to shield from the ambient
UV field the \h2\ clouds of which the \hi\ forms an evaporative
envelope.  They explained that the column density of the \hi\ cloud
around a molecular cloud with no embedded UV sources will not rise above
the value required to self-shield the molecules against dissociation,
due to the global UV field, since any further \hi\ would then convert to
\h2.  We can add here that a similar consideration can be applied to
clouds containing dust, provided that the dust to gas ratio does not
vary dramatically within the disc.  Once the dust-shielding rises above
a critical extinction value any additional \hi\ converts to \h2.  Savage
et al.  (1977) note, from measurements made within the Galaxy, that
conversion to \h2\ sets in for \hi\ column densities of some
$5\times10^{20}$~cm$^{-2}$, corresponding to an $E(B-V)$ of 0.08, which
in turn corresponds to a dust extinction at the dissociation edge for
\h2\ of $\sim$0.7~mag.  Thus the self-shielding explanation of Shaya and
Federman must be supplemented by taking quantitatively into account the
effects of dust, although the result is qualitatively similar. 

We can see in Fig.~4 that the value of the \hi\ column density over the
radial range between 5 and 10~kpc, where it is virtually constant, is a
little less than $5\times10^{20}$~cm$^{-2}$.  At total neutral hydrogen
column densities significantly greater than this all the hydrogen
converts to \h2, except in zones close to local UV sources within the
clouds.  This general tendency is seen as a decline in the \hi\ column
density at radii less than 5~kpc.  Beyond 12~kpc the \hi\ makes up all
the neutral H, and the column density falls exponentially, as do the
surface brightness indices.  It is interesting to note that the arm \hi\
column density in fact stays at its higher level up to $\sim15$~kpc, as
might be expected from the higher concentrations of SF, gas and dust
there.  The difference between our scenario and that of Shaya and
Federman (1987), or the considerations of the physical processes
involved in the \hi\ to \h2\ transition by Elmegreen (1993) is the extra
effect of dust extinction which reduces the \hi\ column required for
shielding, and tends to enhance the \h2/\hi\ ratio throughout the disc. 

Whereas in the inner parts the H{\sc i} may thus be partly produced by
photodissociation of molecular gas by the radiation field from massive
young stars, one cannot assume that the H{\sc i} in the outer regions is
completely independent of the local dissociating radiation field.  Rand
et al.  (1992) find evidence from radial profiles of total gas,
H$\alpha$ and H{\sc i} that dissociation by massive young stars still
plays a role at large radii.  But even if the massive stars are the most
important source of photodissociating radiation, the correlation between
H$\alpha$ and H{\sc i} may break down at large galactocentric radii. 
First, since the total densities of gas and dust are much lower than in
the inner disc, photons can travel further from the stars that produce
them, and molecular hydrogen can be dissociated at greater distances
from the radiation sources, tending to weaken the small-scale H{\sc
i}--H$\alpha$ correlation.  Also, the metallicity may be so low in the
outer regions of the disc that there are not enough dust-grains for
molecular gas to form on.  All these processes can lead to a more
widespread distribution of H{\sc i}.  One example is the tidal arm of
M51 observed in H{\sc i} by Rots et al.  (1990), where no H$\alpha$ is
detected.  Some or all of the above-mentioned processes may cause the
hydrogen in the arm to be predominantly atomic. 

Although the \ha\ emission is used here to trace young massive stars,
this is not entirely adequate for tracing the radiation field which
photodissociates the molecular hydrogen and forms the \hi.  This
radiation field will be more adequately observed in the far UV.  An
important line of future study will be the direct comparison between the
tracers of the supposed origin (UV) and result (\hi) of the
photodissociation.

\section{Conclusions}

We have used images of the barred spiral galaxy NGC~4321 (M100) in
optical ($B$ and $I$ bands), $B-I$ colour, \ha, \hi, and 21cm radio
continuum to study the geometrical distributions of gas and stars in and
outside the spiral arms.  We employ two main techniques.  First, we
overlay the different images to indicate the morphological relations
between massive SF, atomic hydrogen gas, and dust lanes.  Second, we
study azimuthally averaged radial profiles for all tracers, comparing
with CO data, for the whole disc, as well as the arm and interarm zones
separately.  Our main conclusions are the following:

\begin{enumerate}

\item The dust lanes running along most of the spiral arms (and
coinciding with the CO spiral arms in part of them; Rand 1995) are
offset from the \ha\ arms and from the \hi\ arms which coincide
spatially with \ha. This is very similar to the relations observed in
some other spiral galaxies, and can be interpreted as an effect of the
compression of gas by the density wave in the arm, with SF occurring
downstream. \hi\ will be a {\it result} of the massive SF through
photodissociation of part of the molecular gas by the UV radiation field
(cf. Allen et al. 1986).

\item We observe a spatial coincidence of regions with enhanced \ha\
emission (i.e.  with strong current massive SF) and with enhanced \hi\
density throughout the disc of the galaxy.  Most probably the \hi\ in
these regions has its origin in photodissociation by radiation from the
nearby massive stars. 

\item The radial profiles in \ha, blue and NIR light, radio continuum
and also CO have equal (within the errors of the fit) exponential scale
lengths in the region of the SF spiral arms. This is not only true for
the complete disc, but also for the arm, and the interarm regions
separately. The arm scale lengths are however longer (by $\sim50\%$)
than the disc and interarm scale lengths, which are comparable. These
results indicate that the radial profiles are not influenced by dust
extinction or radial  gradients in stellar populations, which would lead
to differences in the scale length values. 

\item The equal disc, arm and interarm scale lengths in \ha, $B$, $I$ and
radio continuum indicate a common origin in star formation for these
profiles. The fact that the CO disc profile has the same scale length
would seem to indicate that the average efficiency of massive SF does
not vary radially in the disc. The CO intensity could however be
proportional to the cosmic ray density as indicated by the radio
continuum profile, in which case the CO intensity would not be a linear
indicator of the density of molecular gas.

\item The \hi\ radial profile is quite different from the \ha, optical,
radio continuum, and CO profiles. After a depression in the central
region (which is not in general confined to barred spirals and cannot be
explained by bar dynamical processes for re-distribution of gas alone)
the profile rises in the region of the SF spiral arms, and is
practically flat in that region. The rise is naturally explained by the
photodissociation process, which should enhance the \hi\ density in the
region of enhanced SF.  The flatness of the profile in the disc is
understandable at least qualitatively in terms of the equilibrium
between atomic and molecular hydrogen in the presence of dust and SF.

\end{enumerate}

{\it Acknowledgements} We thank Drs.  A.H.  Broeils, C.  Carignan and
J.-R.  Roy for comments on an earlier version of the manuscript, and the
anonymous referee for comments that helped improve this paper.

\bsp

\label{lastpage}

\end{document}